\documentclass[sigconf,review=false, nonacm]{acmart}
\usepackage{caption}
\usepackage{subcaption}
\usepackage{dblfloatfix}

\setcopyright{CC}
\setcctype{by-sa}

\begin{document}

\title{Thermal Feedback for Transparency in Human-Robot Interaction}

\author{Svenja Yvonne Schött}
\orcid{0000-0003-1281-0230}
\email{svenja.schoett@ifi.lmu.de}
\affiliation{
\institution{LMU Munich}
\streetaddress{Frauenlobstr. 7a}
\city{Munich}
\country{Germany}
\postcode{80337}
}
\renewcommand{\shortauthors}{Schött}

\begin{abstract}
Robots can support humans in tedious tasks, as well as provide social support. However, the decision-making and behavior of robots is not always clear to the human interaction partner. In this work, we discuss the opportunity of using thermal feedback as an additional modality to create transparent interactions. 
We then present scenarios where thermal feedback is incorporated into the interaction e.g. to unobtrusively communicate the behavior of the robot. We highlight the limitations and challenges of temperature-based feedback, which can be explored in future research.  
\end{abstract}




\maketitle

\section{Introduction}

With the advancement of Human-Robot Interaction (HRI) research, robots move from purely industrial settings to social settings in the home. Instead of product assembly, robots here take care of tasks ranging from cleaning~\cite{fink2013living} and cooking~\cite{oechsner2022challenges} to social interactions. 
Robots that interact with humans via social interactions to provide support, usually in terms of disability and care, are called Socially Assistive Robots (SARs)~\cite{feil2005_defining}. While physical contact is vital for assistive robots who, e.g., help someone off a hospital bed, Feil-Seifer and Matari\a'c~\cite{feil2005_defining} emphasize that physical contact is not mandatory for social support. 
Nonetheless, physical contact with embodied robots is possible and we argue has potential that should be utilized. By adding haptic feedback to a physically embodied robot, additional information can be communicated. Previous work in HRI has already explored vibrotactile feedback~\cite{casalino2018_operator, grushko2021_improved}. 

One of the central challenges in HRI is increasing the transparency in communication between humans and robots. Transparency is a means for a robot to present information with the aim of increasing a human's understanding of the robot's decision-making~\cite{theodorou_designing_2017}. Understanding benefits the interaction, e.g., by establishing appropriate levels of trust in a system~\cite{hastie2017_trust}.
Trust between humans and robots is especially important, as a lack of trust can result in refusal to interact with the robot~\cite{theodorou_designing_2017}.
However, it is still unclear how to establish transparency in communication with SARs, especially focusing on communicating decision-making. 
Due to the physical nature of robots we propose using thermal feedback, a type of haptic feedback~\cite{suhonen2012UX}, to provide supplementary information. We discuss the potential and limitations of thermal feedback for robot-to-human communication. Further, we envision four scenarios where thermal feedback could come into play.

\section{Potential of Thermal Feedback}

The use of thermal feedback has been explored in Human-Computer Interaction (HCI), e.g., to enrich text communication~\cite{wilson2012_thermalicons}.
Thermal feedback is especially well suited for secondary information, as it does not require the user's full attention~\cite{lee2010_thermomessage}. We envision that humans can focus on the actual interaction with the robot while receiving and processing additional information.





Humans have subconscious associations between certain concepts and temperatures. For example, humans may associate light with warmth because a lit candle emits warmth as well~\cite{Lin2013}. These associations go beyond state metaphors but encompass emotions as well. 
Thermal feedback has potential for SARs specifically as it has an emotional component~\cite{ijzerman2009_thermometer}.
Humans have intuitive associations between temperature and emotions based on prior experiences and (western) language~\cite{ijzerman2009_thermometer}. Further, the same brain region is responsible for processing thermal sensations and emotions~\cite{sung2007_brain}. 
For example, we may describe other people as "warm" or "cold" to indicate how affectionate their personalities are. 
These intuitive associations have already been used by HRI researchers. Pe\~{n}a and Tanaka~\cite{pena2020_human} for example changed a robot's temperature and facial expressions to express emotions in accordance with the thermal associations. As the visual stimulus of the facial expressions is stronger, their participants mainly relied on the face of the robot to judge the emotion expression. 

\section{Scenarios for Thermal Feedback in HRI}\label{scenarios}
Based on the discussed characteristics of thermal feedback, we present scenarios of temperature used to enrich human-robot communication. 
\begin{figure*}[!b]
     \centering
     \begin{subfigure}[b]{0.4\textwidth}
         \centering
         \includegraphics[width=0.8\textwidth]{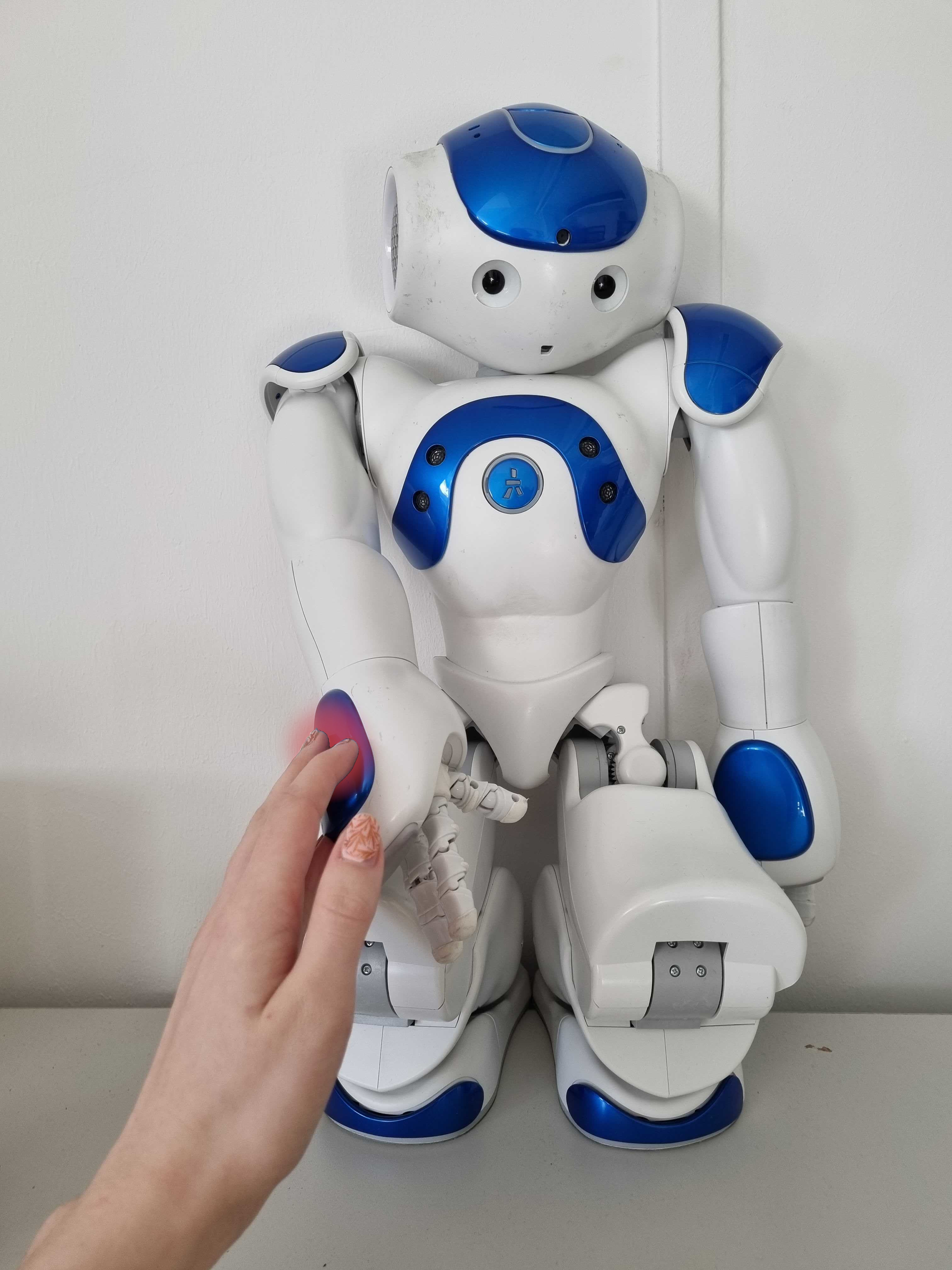}
         \caption{Direct thermal feedback via the robotic arm itself changing temperature.}
         \label{fig:directfeedback}
     \end{subfigure}
\hspace{2em}
     \begin{subfigure}[b]{0.4\textwidth}
         \centering
         \includegraphics[width=0.8\textwidth]{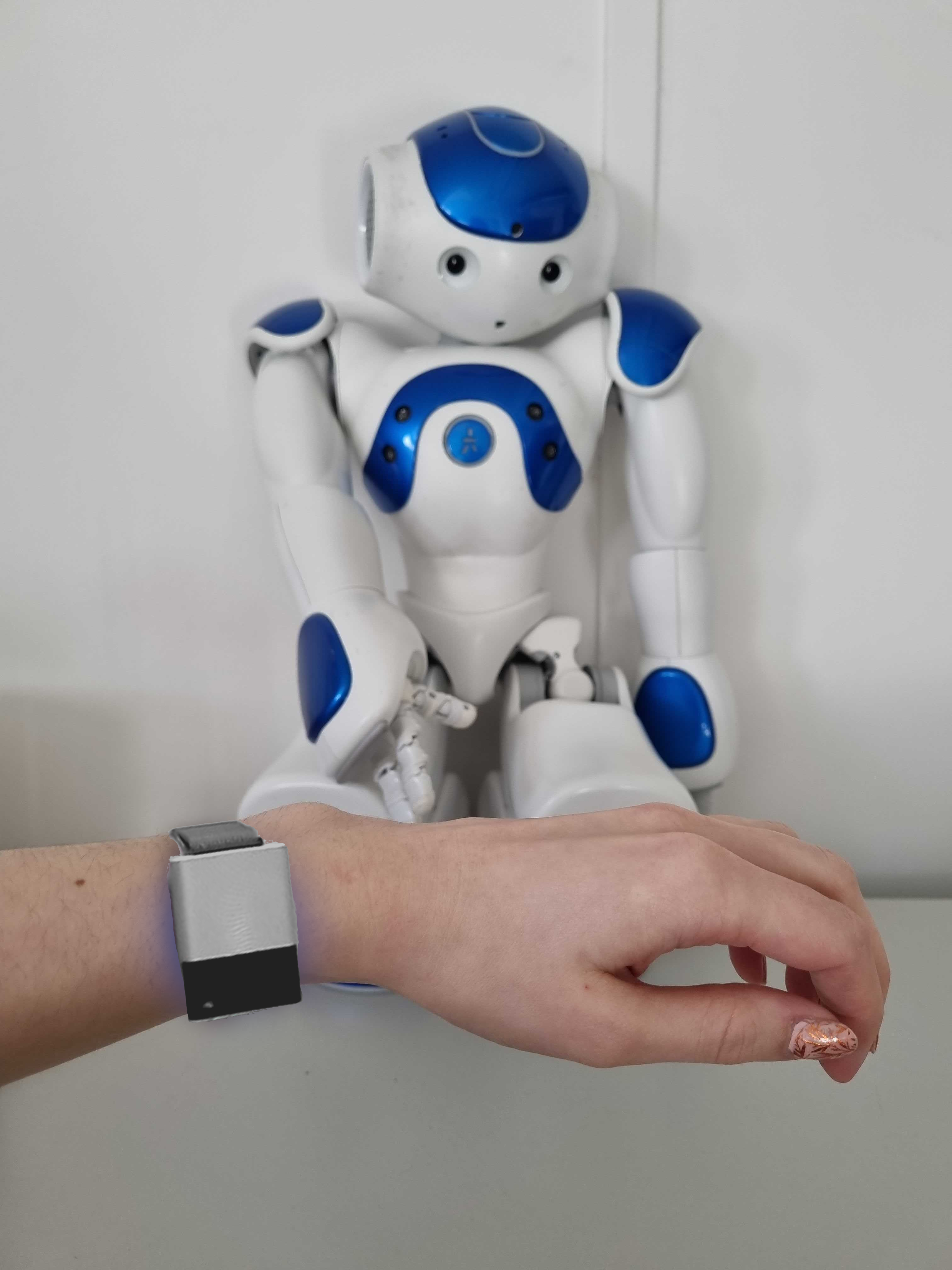}
  \caption{Remote thermal feedback using a thermal feedback bracelet.}
  \label{fig:remotefeedback}
     \end{subfigure}
\caption{Integration of thermal feedback in the interaction with a social robot. }
\label{fig:comparison}
\end{figure*}
\paragraph{Robot Behavior}
Many embodied robots have the ability to move around or manipulate their environment. The most basic approach is to use thermal feedback to communicate this behavior of the robot. 
We envision communicating simple variables in the interaction such as movement, proximity to the human, or the closure of the gripper on a robot arm. For example, there is a working conceptual metaphor for movement, which means that people intuitively associate movement with warmth and stopping with cool temperature~\cite{Lin2013}. Thus, to communicate the next action of a robot of moving somewhere, the robot could turn warmer. Once it gets slower, the temperature cools down.
Here, the thermal feedback can start before the robot executes the behavior to give a "warning", enabling the human interaction partner to anticipate the next steps of the robot and react accordingly. Alternatively, the thermal feedback can accompany the behavior while it is happening. 

\paragraph{Certainty in Decision-Making}
For intelligent robots, it can be hard to discern why they perform a certain action and the certainty in the decision. Valdivia et al.~\cite{valdivia2022_wrapped} explored the communication of certainty of a robot's learning progress with an inflatable haptic display wrapped around a robotic arm. Similarly, a robot arm could turn warm if the robot is "confident" in the next action i.e. that it understands the human correctly. The robot turns cold if the robot is unsure. This temperature change is perceived by a human interaction partner and enables them to provide additional input or react to inappropriate robot actions. 

\paragraph{Social Communication}
From a social perspective, we can envision e.g. an interactive diary robot with which users talk about their feelings. 
Besides communicating specific information to increase transparency in the interaction, the emotional nature of thermal perception can be useful for robots that interact with humans on a social level. Thus, if a robot is engaging in emotional communication, temperature changes could be used to mirror the emotional valence and arousal of the human to show empathy. Empathy expression by SARs has been explored before~\cite{decarolis2017_simulating}. 

\paragraph{Communicating Intentions}
Lastly, we can envision the use of warmth to signify emotional warmth in a social support robot. Instead of using the temperature as a communication tool, it could be part of the robot design, like the texture. The SAR would be slightly warm to communicate the general socially positive intention of support.

\section{Limitations and Challenges}

\subsection{Limited Communication Buffer}
As a primary communication channel, HCI and HRI rely on visual and auditory stimuli. 
Humans can distinguish seven different thermal stimuli~\cite{wilson2015_subjective}. Thus, temperature is limited in its ability to convey complex messages, which sets it apart from visual and auditory modalities.  
Further, as opposed to other commonly employed stimuli, temperature is bipolar. This means that temperature has no natural resting state but two extremes: warmth and cold~\cite{wilson2015_subjective}. By adjusting the rate and degree of temperature change, thermal feedback is best suited for communicating bipolar information. 
Based on these limitations, thermal feedback can not replace visual and auditory feedback, can be useful to supply simple, additional information. 

\subsection{Physiological Limitations}
Thermal feedback has physiological limitations based on the way temperature is perceived.
Firstly, environmental variables like air temperature, humidity~\cite{halvey2012_baby} and presence of clothing between the skin and the thermal module~\cite{halvey2011_clothing} influence the way temperature is perceived. Further, temperature perception is dampened over time i.e. if a human touches something warm for a long time, they become less receptive towards warm feedback~\cite{wilson2012_thermalicons}. Thermal perception is also slower than that of other haptic stimuli~\cite{suhonen2012UX}.
Thermal feedback is therefore only useful for communicating non-urgent information.


\subsection{Thermal Taxonomy}
The central challenge of using thermal feedback for HRI is establishing what concept is being communicated. Going back to the \textit{robot behavior} scenario, movement has a working thermal mapping. If a human knows that the thermal feedback communicates whether a robot is moving, most people have the same concept of what hot vs. cold feedback means. However, what do we do if the human does not know that thermal feedback is communicating movement in the first place? 
The thermal taxonomy may sometimes be clear based on context, but this is not guaranteed. Further research is needed to establish how to best communicate the thermal taxonomy i.e. what the thermal feedback means to a user. 
In addition to this, not all potential information has an intuitive temperature tie. While thermal feedback theoretically can communicate all simple information, this limits where it is easy to implement. 

\subsection{Integrating Thermal Feedback}

In the \textit{communicating intentions} scenario, the robot itself has integrated thermal modules and can become colder and warmer. The human touches their robotic interaction partner, as visualized in~\autoref{fig:directfeedback}. However, most robots do not come with integrated thermal modules to design thermal feedback. 
Alternatively, external thermal displays can be used. In the example in~\autoref{fig:remotefeedback}, the human is wearing a wristband that can change temperatures to receive signals remotely. While additional hardware is required, external thermal displays are useful when it is impractical to constantly touch the robot. For the robot behavior scenario for example an external thermal module is better suited as the robot is moving and the temperature may change more rapidly than for social communication.

\section{Conclusion}
Thermal feedback can communicate missing information to the human during the interaction to increase robot transparency. In this workshop, we present the vision and limitations of thermal feedback for communicating simple and emotional information during HRI. 
On one hand, temperature is perceived subconsciously which leaves room for other channels of communication. On the other hand, only simple information can be communicated. 

Future studies could explore the timing of thermal feedback. Thermal feedback can be transmitted synchronously to a robot action or before the action starts (as a warning). It would be interesting to find out which timing leads to more trust in the robot and transparency. 
More generally, future research could explore what information humans desire for a transparent interaction with robots i.e. what the thermal feedback should communicate. Thereafter, it is necessary to find out if this desired information has usable thermal metaphors. 

We believe that thermal feedback can serve as an additional communication channel to support the interaction. While we argue that thermal feedback has several limitations which need to be considered, we see the potential of adding an additional modality to increase the interaction transparency.




\section{Acknowledgements}
This work was supported by LMUexcellent, funded by the Federal Ministry of Education and Research (BMBF) and the Free State of Bavaria.

\balance
\bibliographystyle{acm-sigchi}
\bibliography{bibliography}

\end{document}